\documentclass[9pt,twocolumn,twoside]{opticajnl}
\journal{opticajournal} % use for journal or Optica Open submissions

\setboolean{shortarticle}{true}
\usepackage{physics}
\usepackage{braket}

\usepackage{lineno}
\title{High-dimensional quantum key distribution using orbital angular momentum of single photons from a colloidal quantum dot at room temperature}

\author[1]{Dotan Halevi}
\author[1]{ Boaz Lubotzky}
\author[1]{Kfir Sulimany}
\author[2]{Eric G. Bowes}
\author[2]{Jennifer A. Hollingsworth}
\author[1]{Yaron Bromberg}
\author[1,*]{Ronen Rapaport}

\affil[1]{Racah Institute of Physics, The Hebrew University of Jerusalem, Jerusalem 9190401,Israel}
\affil[2]{Materials Physics \& Applications Division: Center for Integrated Nanotechnologies, Los Alamos National Laboratory,
Los Alamos, New Mexico 87545, USA}
\affil[3]{School of Optics, University of Technology, 2000 J St. NW, Washington DC, 20036}

\affil[*]{ronen.rapaport@mail.huji.ac.il}

\begin{abstract}
High-dimensional quantum key distribution (HDQKD) is a promising avenue to address the inherent limitations of basic QKD protocols. However, experimental realizations of HDQKD to date have relied on indeterministic photon sources that limit the achievable key rate. In this paper, we demonstrate a full emulation of a HDQKD system using a single colloidal giant quantum dot (gQD) as a deterministic, compact and room-temperature single-photon source (SPS). We demonstrate a practical protocol by encoding information in a high-dimensional space ($d = 3$) of the orbital angular momentum of the photons. Our experimental configuration incorporates two spatial light modulators for encoding and decoding the spatial information carried by individual photons. Our experimental demonstration establishes the feasibility of utilizing high radiative quantum yield gQDs as practical SPSs for HDQKD. We also demonstrate experimentally secure qudit transmission exceeding one secure bit per photon, thus already beating the traditional $d=2$ QKD capacity.
\end{abstract}

\setboolean{displaycopyright}{false} % Do not include copyright or licensing information in submission.

\begin{document}

\maketitle

\section{Introduction}
High-dimensional quantum key distribution (HDQKD) offers an appealing approach to enhancing the performance of basic QKD systems. Unlike traditional QKD protocols, which are based on encoding two-dimensional quantum bits (qubits) using two optical modes, such as linear polarization states, HDQKD leverages $d>2$ modes to encode d-dimensional quantum bits (qudits) with a single photon. The higher information capacity significantly increases the secure key rate per photon and also enhances the protocol's resilience to quantum bit error rate \cite{bruss1998optimal, cerf2002security, bechmann2000quantum}. HDQKD protocols have been demonstrated utilizing spatial \cite{walborn2006quantum,etcheverry2013quantum,mirhosseini2015high,ding2017high,sit2017high,bouchard2017high,bouchard2018experimental,cozzolino2019orbital,tentrup2019large,zhou2019using,otte2020high,da2021path,hu2021pathways,ortega2021experimental,stasiuk2023high,lib2024high}, time-bin\cite{islam2017provably,lee2019large,vagniluca2020efficient,ikuta2022scalable,chapman2022hyperentangled,sulimany2021fast}, or time-energy \cite{ali2007large,mower2013high,lee2014entanglement, zhong2015photon,liu2019energy,bouchard2021achieving,liu2023high,bulla2023nonlocal,chang2023large} encoding.

The spatial mode of a photon is a convenient degree of freedom for encoding qudits, since it is relatively easy to manipulate the states of such a qudit using phase plates and spatial light modulators (SLMs). One notable spatial mode basis is that of the orbital angular momentum (OAM) of photons \cite{yao2011orbital}. OAM modes are characterized by a helical phase of the electric field, given by $e^{i\phi l}$, where $\phi$ is the azimuthal angle and $l=0,\pm 1,\pm 2...$ is the quantum orbital angular momentum number. Thus, each OAM mode carries a specific orbital angular momentum value and is orthogonal to the other modes. These modes can be used as a basis for encoding information.

%OAM bases  provide an optimal balance between  complexity and distinguishability, offering a practical compromise that enhances the information-carrying capacity while facilitating experimental implementation. This approach capitalizes on the spatial degree of freedom, and devices such as light plates and SLMs serve as encoders and decoders for the spatial modes \cite{sit2017high, vallone2014free, nape2023quantum, mafu2013higher}.

Similar to traditional two-dimensional QKD protocols, ideal HDQKD implementations require a true single-photon source (SPS). However, until now, experiments demonstrating HDQKD relied only on alternative sources: (a) attenuated laser pulses \cite{vallone2014free, mirhosseini2015high, liao2017satellite}, which generate weak coherent photon states, and are generally vulnerable to photon number splitting attacks, which limits the maximal secure key rate and distance \cite{lutkenhaus2002quantum,scarani2004quantum,acin2004coherent,wang2005beating}, or 
(b) sources based on spontaneous parametric down-conversion (SPDC) \cite{nape2023quantum, yesharim2023direct, mafu2013higher, sit2017high}, which generate entangled photon pairs but suffer from indeterminism in the photon production times and low photon-pair emission probabilities. These photon sources, therefore, limit the performance of practical HDQKD systems. For this reason, a demonstration of HDQKD using an SPS is essential. 

An emerging SPS is the CdSe/CdS core/thick-shell giant colloidal quantum dot (gQD) \cite{chen2008giant, ghosh2012new}.
%Recognizing these limitations is crucial for refining techniques and advancing robust quantum communication protocols.
Recent advancements in gQD synthesis\cite{orfield2018photophysics} and integration \cite{dawood2018role} into nanoantenna devices yielded superior performance of gQDs as promising sources for single photons. gQDs coupled to nanoantennas have several advantages, including room-temperature operation and stability over time \cite{chen2008giant, dawood2018role, ghosh2012new}, high single-photon purity \cite{abudayyeh2019purification}, very fast emission rates with high photon directionality leading to high brightness \cite{abudayyeh2021overcoming} and near unity collection efficiency \cite{abudayyeh2021single}. Additionally, their ease of synthesis, versatility of integration with various nanostructures, and tunable emission wavelengths \cite{krishnamurthy2020pbs, dennis2019role} make them attractive candidates for large-scale quantum communication networks. This paves the way for gQDs as a practical SPS for QKD systems, particularly in HDQKD. However,  demonstration of encoding and measuring HDQKD bases using single photons from gQDs or other room-temperature SPSs is still an outstanding challenge.

 Herein, we successfully emulate a free-space HDQKD system and protocol based on two $d=3$ mutually unbiased bases (MUBs) of OAM states, encoded on (Alice) and decoded from (Bob), single photons emitted from a single gQD. The encoding and decoding are done using two SLMs. We demonstrate experimentally secure qudit transmission exceeding one secure bit per photon, thus already surpassing the traditional two-dimensional QKD capacity.

\section{Experimental concept}

In order to implement an HDQKD protocol using OAM imprinted on single photons, we choose two MUBs based on OAM states, each with dimension $d=3$. The first basis, MUB$_1$, consists of three states $\{\ket{i}\}=\{ \ket{a},\ket{b},\ket{c} \}$ with an azimuthal phase having quantum numbers $l=-1,0,1$, respectively, and a Gaussian intensity profile. The second basis, MUB$_2$, consists of three mutually orthogonal states, $\{\ket{j}\}=\{ \ket{\alpha},\ket{\beta},\ket{\gamma}\}$, each is a linear combination of the states of the first basis, chosen such that $|\braket{i|j}|^2=\frac{1}{3}$, as is detailed in table \ref{demo-table}. 

\subsection{Experimental setup and SPS characterization}

\begin{figure*}[t!]
\centering\includegraphics[width=13cm]{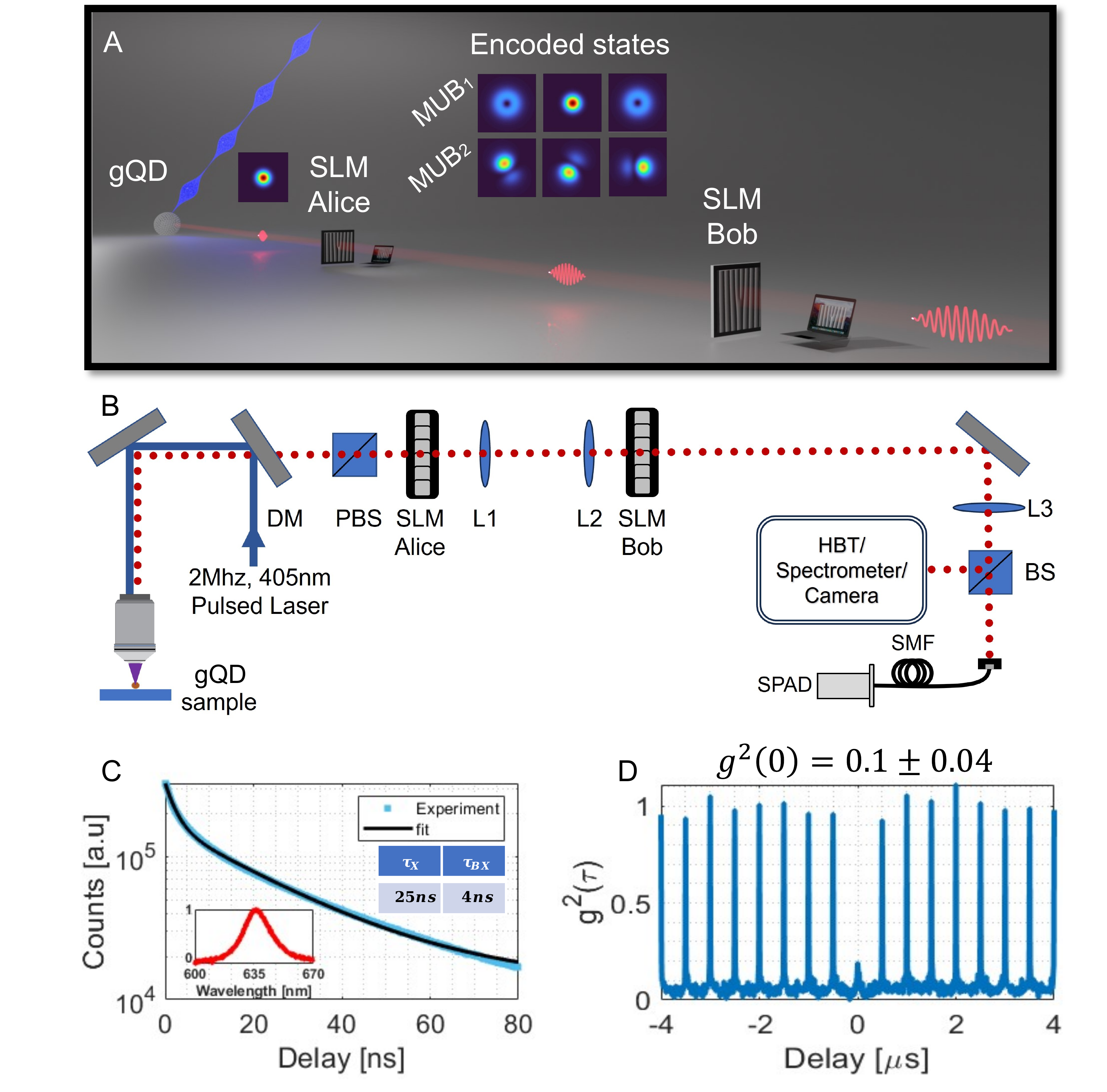}
\caption{\textbf{Demonstration of high-dimensional quantum key distribution using giant quantum dot (gQD) based single photon source (SPS): Experimental concept and setup, SPS characterization.} 
    (A) Visualization illustrating the key
    steps in the experiment: Starting from the left, a pulsed laser excites the gQD, which in turn emits single photons with a spatial Gaussian profile (as indicated by the image above the photon). These photons accumulate a spatial phase pattern by Alice’s SLM, encoding spatial information in one of six modes selected randomly by Alice ("encoded states” as seen in the far field). The information is then transferred to Bob’s SLM, which is used to project the photon on of the same size modes, selected randomly by Bob.
    (B) Schematic of the experimental setup. Moving from left to right, a gQD sample on glass, a high NA objective (Olympus MPLFLN100xBD), dichroic mirror (DM) used to reflect the 2Mhz repetition rate, 405nm, 1ns pulsed blue laser towards the gQD while allowing only the emitted red photons to pass, a PBS to polarize the emitted photons, Alice's SLM, a 4f optical system ($f_{L1}=f_{L2}=20cm)$, Bob's SLM, plane lens ($f_{L3}=30cm$), and a BS. The BS directs photons to either an SMF for projection measurement or to an analysis setup: HBT, spectrometer, or camera for further analysis. 
    (C) The blue curve presents the measured emission lifetime of the single gQD, the black line corresponds to a fitted 2-term exponential, accounting for exciton ($\tau_x$) and biexciton ($\tau_{Bx}$) lifetimes of 25ns and 4ns, respectively. The inset displays the spectrum of the gQD, centered at $640$ and with a FWHM of $30nm$.   
    (D) The measured second order correlation function, $g^{(2)}(\tau)$ of the single gQD, displaying $g^{(2)}(\tau=0)=0.1\pm 0.04$ after time gating of $11ns$. } 
    \label{fig:3dCQD}
\end{figure*}

We start by characterizing the gQD-based SPS. Fig. \ref{fig:3dCQD}C shows the measured emission decay and spectrum of the single gQD (on a glass substrate) under pulsed laser excitation (405nm, 2MHz). The radiative time trace is well fitted to a bi-exponent decay, corresponding to the biexciton-exciton cascaded emission. %, , we measure a photon rate of approximately 8000 photons per second from a single gQD. Our system suffers from a loss of 10\%. The emitted photons has a central wavelength of $640nm$ and a spectral bandwidth of $15nm$ at FWHM (fig \ref{fig:3dCQD}C).
%In terms of quantum characteristics, the gQD's emission is governed by excitonic and biexcitonic processes. The exciton, formed by an electron-hole pair, has a lifetime of approximately $25ns$, while the biexciton, involving two such pairs, has a shorter lifetime of around $4ns$ (fig\ref{fig:3dCQD}B). 
To mitigate the effects of more than one photon per pulse due to a radiative biexciton-exciton emission, a temporal filtering techniques has been applied to the the gQD SPS \cite{abudayyeh2021overcoming,abudayyeh2021single}
improving the single photon purity, as is evident in second-order correlation measurement presented in Fig. \ref{fig:3dCQD}D. After filtering of $11ns$ we get $g^{(2)}(0)<0.1\pm 0.04$ including the correlated noise of the detector. It is important to note that this already high single photon purity of gQDs at room temperatures can be greatly improved by applying simple photon purification methods, with a demonstrated photon purity $P_1>0.995$, which means a two-photon probability $P_{>1}<0.5\%$, limited by detector dark counts   \cite{abudayyeh2019purification}. This very low value of $P_{>1}$ emphasizes the suitability of CdSe/CdS gQDs as excellent SPSs for quantum communication applications.

\begin{table}
\begin{center} 
\begin{tabular}{||c c||} 
 \hline
 $MUB_1$ & $MUB_2$ \\ [0.5ex] 
 \hline\hline
 $\ket{a}=OAM_{l=-1}$ & $\ket{\alpha}=\frac{1}{\sqrt{3}}(\ket{a}+\ket{b}+z^2\ket{c})$ \\ 
 \hline
 $\ket{b}=OAM_{l=0}$ &  $\; \ket{\beta}=\frac{1}{\sqrt{3}}(\ket{a}+z\ket{b}+z\ket{c}$) \\ 
 \hline
  $\ket{c}=OAM_{l=+1}$ &  $\ket{\gamma}=\frac{1}{\sqrt{3}}(\ket{a}+z^2\ket{b}+\ket{c})$ \\ 
 \hline
\end{tabular}
\caption{\label{demo-table} \textbf{Two d=3 OAM-based mutually unbiased Bases for HDQKD.} $\{ \ket{a},\ket{b},\ket{c} \}$ represent basis states in $MUB_1$ and $\{ \ket{\alpha},\ket{\beta},\ket{\gamma}\}$ represent basis states in $MUB_2$. Here $z = exp(2\pi i/3)$.}
\end{center}
\label{fig:MUBs}
\end{table}

%\begin{figure}[H]
%    \includegraphics[width=0.5\textwidth]{liferime_g2.jpg}
%    \centering
%    \caption{\textbf{Characterization of gQD.} 
%   (A) Lifetime Curve and Spectrum. In the primary subfigure, the blue curve represents lifetime measurements of our gQD, and the black line corresponds to a fitted 2-term exponential, accounting for exciton ($\tau_x$) and biexciton ($\tau_{xx}$) lifetimes of 25ns and 4ns, respectively. The inset displays the spectrum of the gQD, centered at $640\pm 15nm$.
%   (B) Photon Statistics. Shows the $g^{(2)}$ measurement, revealing  $g^{(2)}(\tau=0)=0.1\pm 0.04$. This metric, where  $g^{(2)}(\tau=0)<0.5$ ensures single-photon behavior, indicate the single-photon purity of our gQD.} 
%    \label{fig:CQD}
%\end{figure}

%To showcase this concept, we conducted an experiment utilizing our gQDs, as single-photon sources (fig \ref{fig:Setup}). The gQDs were spin-coated onto a glass substrate, providing a controlled and uniform distribution. To excite the gQDs, a laser pump operating at $405nm$ with a pulse rate of $2MHz$ from a Toptica laser was employed. The gQDs, upon excitation, emitted photons at $645\pm 15nm$, serving as point sources for our quantum communication system.

A schematic description of HDQKD emulation experiment is shown in Fig. \ref{fig:3dCQD}B. The photons emitted from the gQD were collected using a high numerical aperture (NA 0.9) objective lens (Olympus MPLFLN100xBD). Subsequently, the reflected laser radiation was filtered out by a dichroic mirror (DM), and the collected photons were then linearly polarized using a polarizing beam splitter (PBS).

The first SLM (Alice) (phase-only Holoeye Pluto - NIR-011), was then programmed to encode any of the six states in table \ref{demo-table} on the photon's spatial mode. The states of MUB$_1$ are directly encoded using the phase-only SLM by applying the required helical phase mask. Since the states of MUB$_2$ are suppositions of OAM modes, their encoding requires amplitude and phase modulation. Nonetheless, we approximate these states by states with the same phase profile and a Gaussian amplitude  (see SM 1A and 1B). 

A 4f imaging setup was employed to image SLM-Alice onto a second SLM-Bob (with a total distance of $80 cm$ between the two SLMs), allowing for the seamless transmission of the encoded information. SLM-Bob was programmed to decode the information by applying the complex-conjugated phases of the states of MUB$_1$ and the phase-only approximated states of MUB$_2$ \cite{qassim2014limitations}. The overlap of the states encoded by Alice and decoded by Bob is proportional to the probability of detecting the photon on the optical axis at the far-field of the SLMs. It was measured by placing single-mode fiber coupled to a single photon avalanche photodiode (SPAD) at the focal plane of a lens (L3) placed after SLM-B. A non-polarizing beam splitter (BS), placed between the lens and the fiber, probabilistically directs some of the photons to an optical analysis setup consisting of a Hanbury Brown and Twiss (HBT) setup, a spectrometer, and sCMOS camera, all placed at the same plane as the single mode fiber. %This versatile setup enabled us to explore both the quantum properties of the photons and their potential applications in quantum key distribution.

\section{Results}
\subsection{Imaging of the spatial mode of photons after decoding}

\begin{figure}[htbp]
\centering\includegraphics[width=7cm]{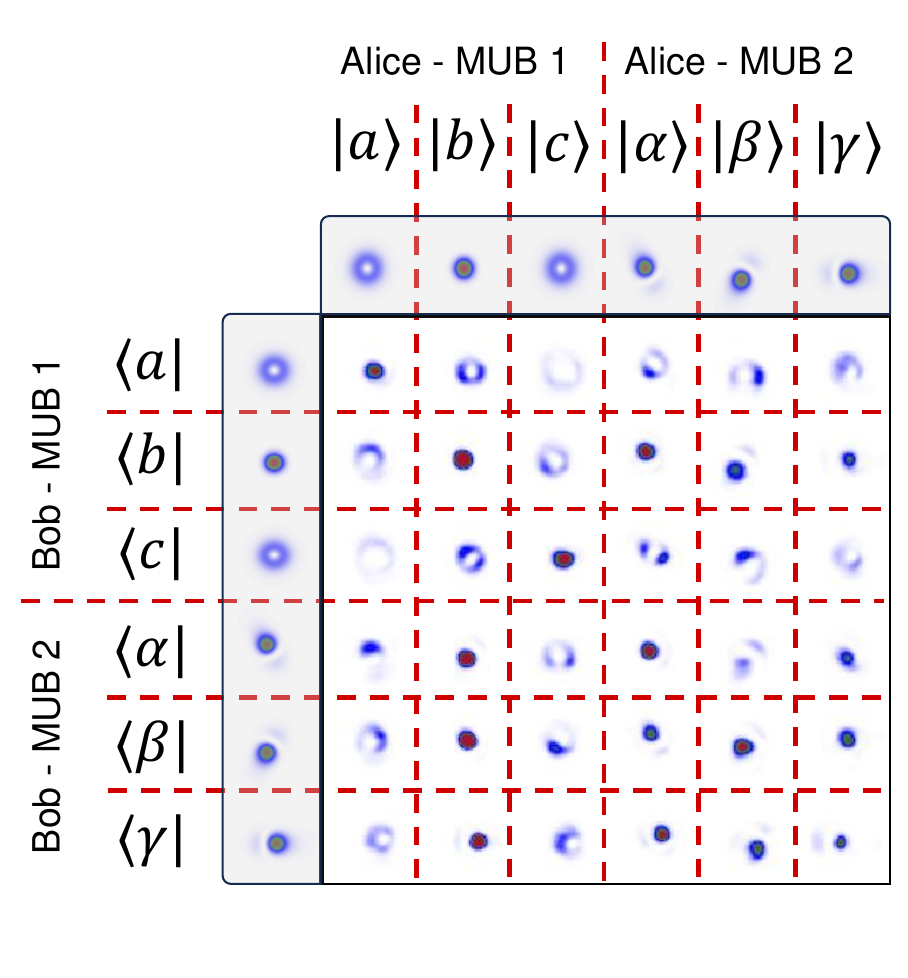}
\caption{\textbf{Imaging the spatial profile of decoded photons in a d=3 OAM-based encoding on two mutually unbiased bases (MUBs.)} 
  This matrix captures encoded photons by Alice and decoded by Bob, taken by a camera. Each row represents a particular mode chosen by Bob, and each column indicates the mode selected by Alice. The main diagonal displays near-perfect Gaussian beams, signifying matched bases for optimal projection into the single-mode fiber. Surrounding the matrix, a complete row/column illustrates simulated modes from Table 1.}
  \label{fig:crosstalks}
\end{figure}

The first part of the emulation experiment involved imaging the photons encoded by Alice and decoded by Bob using a high-resolution camera. Each image is acquired by a long exposure of 60 seconds. The resulting matrix, with the columns representing the modes selected by Alice and the rows are modes Alice, is presented in (fig Fig. \ref{fig:crosstalks}). The modes displayed at the perimeter of the image are the calculated modes of Table 1.

As can be clearly seen, the main diagonal of the matrix results in an almost perfect Gaussian mode at the center of the image, indicating a situation where Alice and Bob decided to encode and decode in the same basis and the same mode. Importantly, considering imaging of the decoded photons onto an SMF, the diagonal elements predict an optimal coupling efficiency. The off-diagonal elements in the two diagonal blocks, all have near-zero intensity at their center. Therefore imaging this mode onto an SMF, will result in minimal, ideally zero, coupling efficiency. This emulates the situation where Bob selected the correct basis but the wrong element. The off-axis blocks, which represent the cases where Bob and Alice have selected different MUBs, are hard to interpret just by looking at the image. A clearer understanding can be gained from an actual projection of the resulting modes onto an SMF (Section \ref{Projection measurement into a single mode fiber})

\subsection{Projection measurement into a single mode fiber}
\label{Projection measurement into a single mode fiber}
\begin{figure}[htbp]
\centering\includegraphics[width=8cm]{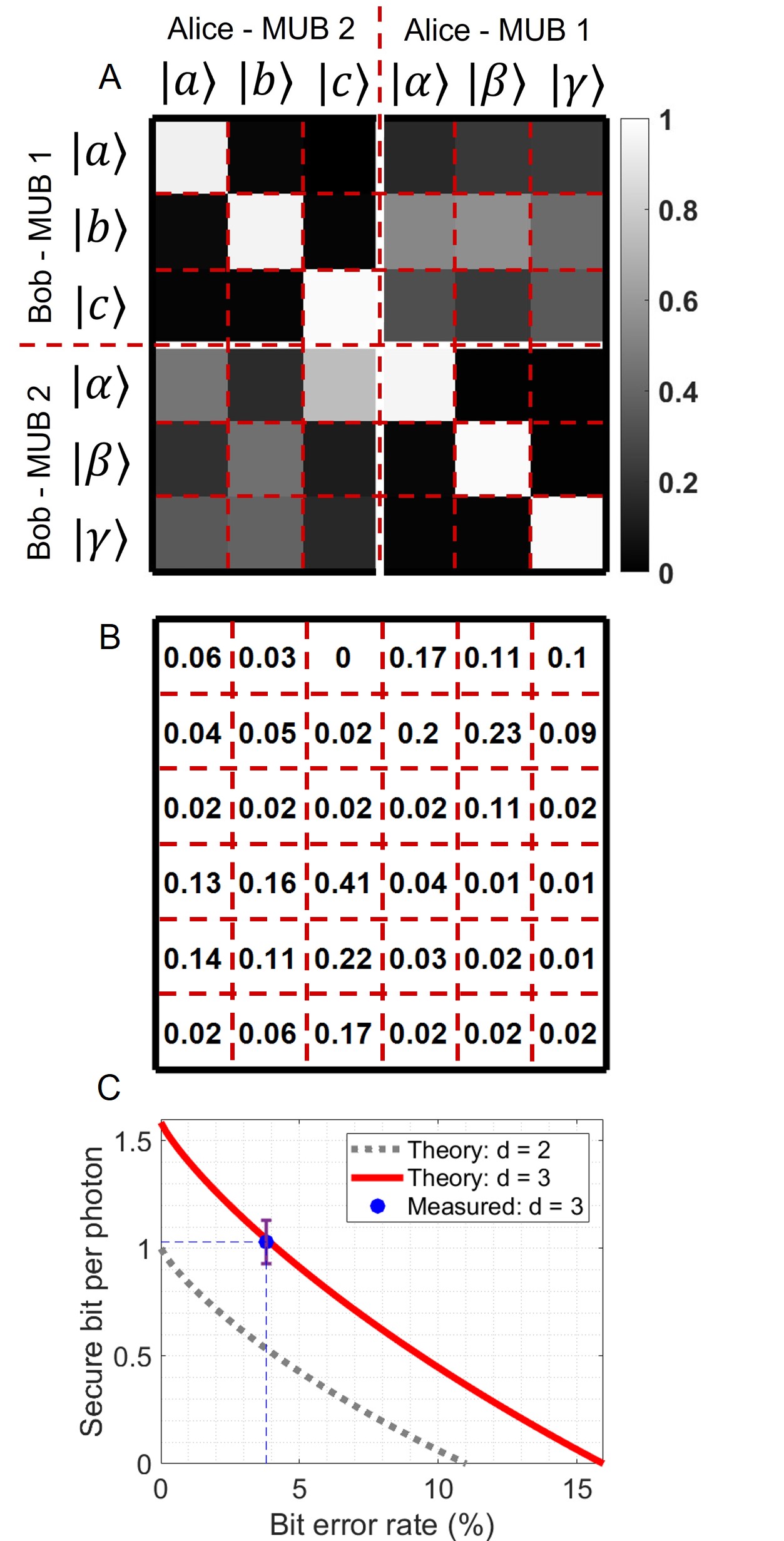}
\caption{\textbf{Emulation of a d=3 high-dimensional quantum key distribution protocol using single photons.} 
  (A) The normalized counts recorded at the single photon detectors, when Alice and Bob each select one of the three states from one of the two bases. (B) The deviation of (A) from theory (C) The secret key rate is plotted against the average error rate $e_b$ for dimensions $d=2$ and $d=3$. The solid data point represents the measured error rate in our system. The solid curve corresponds to the theoretical values for $d=3$, while the dashed curve represents the theoretical values for $d=2$. We get a bit error rate of $3.8\%$ and thus from Eq.~(\ref{Rate}), a secure bit per photon of $1.03$.} 
    \label{fig:projections}
\end{figure}

To demonstrate a full HDQKD decoding system at Bob's side, we conducted single photon projection measurements by imaging the decoded photons after SLM-Bob onto an SMF connected to a SPAD. 
The results of these measurements are presented in Fig. \ref{fig:projections}A. Here we present the SPAD counts at each configuration of SLM-Alice and SLM-Bob, normalized such that each column in each 3x3 block sums to 1.
Consistent with our camera imaging results, the diagonal elements exhibit a nearly perfect projection of $96.2\% \pm 1.3\%$, while a very low projection of $1.9\% \pm 0.9\%$ is measured in all the off-diagonal elements. In the off-diagonal blocks, a nearly uniform projection of around $1/3 \pm 0.16$ is observed, which is expected for measuring in a MUB  with $d=3$, as explained above: a measurement in the wrong basis should yield a projection of $\frac{1}{3}$, thus giving no information.

Next, based on the measurement results, we extract the projected secure key rate of our HDQKD emulation experiment. The secure key rate per photon is a critical metric for evaluating the efficiency of any QKD system. It quantifies the amount of secure key generated per detected photon, thus quantifying the performance of the QKD system.
Theoretically, the rate at which secret key bits are generated per sifted photon, denoted as the secure key rate $R$, is defined as \cite{sheridan2010security}:

\begin{equation}
    R = log_2(d)-h^{(d)}(e_{b1}) -h^{(d)}(e_{b2})
    \label{Rate}
\end{equation}
where $e_{1}$ and $e_{2}$ are the quantum bit error rates of MUB1 and MUB2 respectively and $h^{(d)}(x) = -xlog_2(x/(d-1))-(1-x)log_2(1-x)$ is the $d$-dimensional Shannon entropy.
In traditional QKD systems with $d=2$, the secure key rate per photon is inherently limited to $R\leq 1$. However, as can be seen from Eq. \ref{Rate}, a HDQKD with $d=3$ can reach $R=1.58$ in a noiseless system.
The theoretical prediction for $R$ as a function of the bit error rate for $d=2$ and $d=3$ are plotted by the dotted and full lines in Fig. \ref{fig:projections}B, respectively.

 To calculate the bit error rate of our system, we compute the average deviation from theoretical expectations across the diagonal blocks of Fig. \ref{fig:projections}A, which assumes a constant noise, since in our system it arises mostly from the dark noise of the detectors. This analysis yields a bit error rate of $3.6\% \pm 1.6\%$ for MUB1 and $4.0 \% \pm 0.8\%$ for MUB2. Using Eq.~(\ref{Rate}), we find $R = 1.0 \pm 0.1$. This is a dramatic improvement over the achievable secure bit rate of $d=2$ QKD system having the same quantum error rate (Fig: \ref{fig:projections}B). We note that $d>2$ is advantageous even in the presence of large system noise, as the maximal tolerated noise is significantly higher, 15.8\% compared with 11\%, as seen from Fig \ref{fig:projections}B. This underscores the superior performance and robustness of HDQKD protocols in the face of noise challenges. We note that unlike bit error rate extracted form the diagonal blocks, the errors in the off-diagonal blocks arises mostly from the imperfection in the decoded modes resulting from our phase-only approximation (Fig. S1).

\section{Discussion}
Our experimental demonstration establishes the feasibility of utilizing a high radiative quantum yield, room-temperature gQD  as a compact SPS for high-dimensional quantum key distribution. Here, we used a bare gQD, which is limited in its photon emission rate due to its inherent radiative lifetime, and in the photon collection efficiency, due to the isotropic emission pattern. In recent years, we have shown that both these limitations can be greatly improved by coupling these gQDs to a hybrid metal-dielectric antenna consisting of a nanocone resonator surrounded by a circular Bragg bullseye antenna \cite{abudayyeh2021overcoming,abudayyeh2021single,lubotzky2024room, nazarov2024ultrafast}. These SPS devices showed both record high photon directionality and collection efficiency, together with a sub-nanosecond emission lifetime, enabling a GHz rate of single photons from a room-temperature source. With the recently demonstrated single-photon purification technique yielding negligible two-photon events \cite{abudayyeh2019purification}, these ultrabright sources combined with the HDQKD scheme shown here will result in much higher bit rates and much lower quantum bit error rates leading to new and exciting opportunities for robust, compact and fast quantum encryption systems based on deterministic photon sources. Furthermore, an added ability to position several gQDs with different emission wavelengths on the same nano-antenna could enable encoding HD qudits simultaneously on distinguishable photons, which can be used for distributing a common secret key in a quantum network from a single SPS device.

To fully leverage the potential of gDS for generating single photons at a GHz rate for HDQKD, faster encoding and decoding schemes are essential. High-dimensional encoding can be achieved using a fast optical switch to route photons through static phase plates \cite{cozzolino2019orbital,graham2023multiscale}, while efficient decoding can be achieved with mode-sorters \cite{lavery2012refractive,mirhosseini2013efficient,lightman2017miniature,fickler2020full} or multi-plane light converters \cite{hiekkamaki2021high,lib2022processing}, focusing each OAM and MUB state to a separate single photon detector. Integrating gDS with these technologies opens the door to realizing QKD systems with unprecedented performance over noisy links.

\begin{backmatter}
\bmsection{Funding}

This research was supported by the Quantum Communication Consortium of the Israeli Innovation Authority and the Zuckerman STEM Leadership Program.
\bmsection{Acknowledgments}
The gQD synthesis was performed at the Center for Integrated Nanotechnologies (CINT), a Nanoscale Science Research Center and User Facility operated for the U.S. Department of Energy (DOE) Office of Science. E.G.B. was funded by CINT. J.A.H. acknowledges funding through the U.S. DOE, Office of Science, Office of Advanced Scientific Computing Research, Quantum Internet to Accelerate Scientific Discovery Program. 

\bmsection{Disclosures}
The authors declare no conflicts of interest.

\bmsection{Data Availability}
Data underlying the results presented in this paper are not publicly available at this time but may be obtained from the authors upon reasonable request. 

\end{backmatter}

% Bibliography
\bibliography{sample}

% Full bibliography added automatically for Optics Letters submissions; the following line will simply be ignored if submitting to other journals.
% Note that this extra page will not count against page length
\bibliographyfullrefs{sample}

%Manual citation list
%\begin{thebibliography}{1}
%\bibitem{Zhang:14}
%Y.~Zhang, S.~Qiao, L.~Sun, Q.~W. Shi, W.~Huang, %L.~Li, and Z.~Yang,
 % \enquote{Photoinduced active terahertz metamaterials with nanostructured
  %vanadium dioxide film deposited by sol-gel method,} Opt. Express \textbf{22},
  %11070--11078 (2014).
%\end{thebibliography}

\end{document}